\documentstyle[12pt,bezier]{article}
\begin{document}
\def \inbar{\vrule height1.5ex width.4pt depth0pt}
\def \xC{\relax\hbox{\kern.25em$\inbar\kern-.3em{\rm C}$}}
\def \xR{\relax{\rm I\kern-.18em R}}
\newcommand{\xZ}{Z \hspace{-.08in}Z}
\newcommand{\xbe}{\begin{equation}}
\newcommand{\be}{\begin{equation}}
\newcommand{\xee}{\end{equation}}
\newcommand{\ee}{\end{equation}}
\newcommand{\xbea}{\begin{eqnarray}}
\newcommand{\bea}{\begin{eqnarray}}
\newcommand{\xeea}{\end{eqnarray}}
\newcommand{\eea}{\end{eqnarray}}
\newcommand{\xnn}{\nonumber}
\newcommand{\nn}{\nonumber}
\newcommand{\xkt}{\rangle}
\newcommand{\kt}{\rangle}
\newcommand{\xbr}{\langle}
\newcommand{\br}{\langle}
\newcommand{\xcun}{\mbox{\footnotesize${\cal N}$}}
\newcommand{\cun}{\mbox{\footnotesize${\cal N}$}}
\title{Noncyclic Geometric Phase and Its Non-Abelian Generalization}
\author{Ali Mostafazadeh\thanks{E-mail address: 
amostafazadeh@ku.edu.tr}\\ \\
Department of Mathematics, Ko\c{c} University,\\
Istinye 80860, Istanbul, TURKEY}
\date{ }
\maketitle

\begin{abstract}
We use the theory of dynamical invariants to yield a simple derivation 
of  noncyclic analogues of the Abelian and non-Abelian geometric 
phases. This derivation relies only on the principle of gauge invariance 
and elucidates the existing definitions of the Abelian noncyclic geometric 
phase. We also discuss the adiabatic limit of the noncyclic geometric 
phase and compute the adiabatic non-Abelian noncyclic geometric phase
for a spin 1 magnetic (or electric) quadrupole interacting with a precessing 
magnetic (electric) field.
\end{abstract}
\vspace{2mm}

\vspace{2mm}

\baselineskip=24pt

\section{Introduction}

Since the publication of Berry's seminal paper \cite{berry-1984} on the
adiabatic geometric phase, the concept of geometric phase has been
generalized in a number of ways. Following the work of Aharonov 
and Anandan \cite{a-a} on nonadiabatic geometric phase, Samuel and Bhandari 
\cite{sa-bh} showed that one can indeed define an analogue of the Abelian geometric 
phase for a quantum state that does not undergo a cyclic evolution. Zak \cite{zak}, 
Aitchison and Wanelik \cite{ai-wa}, Mukunda and Simon \cite{mu-si}, 
Pati \cite{pati,pati2}, and de~Polavieja and Sj\"oqvist \cite{po-sj} have ellaborated on the 
theoretical aspects of noncyclic geometric phases, and Wu and Li \cite{wu-li}, 
Weinfurter and Badurek \cite{we-ba}, Christian and Shimony \cite{ch-sh}, Wagh and 
Rakhecha \cite{wa-ra} and Wagh, Rakhecha, Fischer, and Ioffe \cite{wa-ra-fi-io} have 
explored its experimental consequences. In all these investigations the authors consider
Abelian noncyclic geometric phases. The main purpose of the present paper is to 
offer an alternative approach to noncyclic geometric phases which clarifies the 
existing results on the Abelian noncyclic geometric phase and allows for its 
non-Abelian generalization.

The cyclic geometric phase can be conveniently discussed within the framework of the 
theory of dynamical invariants of Lewis and Riesenfeld \cite{le-ri}. The application of 
dynamical invariants in the study of the cyclic geometric phases has been considered by 
Morales \cite{morales} and Monteoliva, Korsch and N\'u$\tilde{\rm n}$es \cite{mkn}
for the Abelian case and by the present author \cite{jpa-98c} for the non-Abelian case. 

In this article, we shall first present a brief review of the necessary results from 
the theory of dynamical invariants and comment on their relevance to the cyclic 
geometric phases in section~2. In section~3, we outline an alternative approach to 
nonadiabatic cyclic geometric phase which is essentially the same as Berry's approach 
to the adiabatic geometric phase  \cite{berry-1984}. This generalizes the analysis of 
Ref.~\cite{jpa-93} in which such an approach is developed for the description of the 
nonadiabatic cyclic geometric phase of a magnetic dipole interacting with a precessing 
magnetic field. In section~4, we derive an expression for the evolution operator and 
discuss its gauge invariance. In section~5, we give our definition of the noncyclic 
geometric phase and explore its relationship to the cyclic geometric phase 
\cite{anandan,jpa-98c}. In section~6, we restrict to the Abelian case and compare our 
definition of the noncyclic Abelian geometric phase with the earlier definitions 
\cite{ai-wa}. In section~7, we present a discussion of the adiabatic approximation and 
show that our analysis reproduces the results of  Ref.~\cite{po-sj} in the adiabatic limit.
In section~8, we calculate the adiabatic non-Abelian noncyclic geometric phase for
a spin 1 magnetic (or electric) quadrupole interacting with a precessing magnetic 
(resp.\ electric) field. Finally, we present our concluding remarks in section~9.

\section{Invariant Operators and Cyclic Geometric Phase}

As shown by Lewis and Riesenfeld \cite{le-ri}, one can write the solution 
of the Schr\"odinger equation
	\xbe
	i\frac{d}{dt}|\psi(t)\kt=H(t)|\psi(t)\kt
	\label{sch-eq}
	\xee
for a time-dependent Hamiltonian $H(t)$, as a linear combination of certain
eigenvectors of a Hermitian dynamical invariant. The latter is a
Hermitian operator $I(t)$ satisfying
	\xbe
	\frac{dI(t)}{dt}=i\left[ I(t),H(t)\right]\;.
	\label{dyn-inv}
	\xee
We shall assume that both $H(t)$ and $I(t)$ have discrete spectra. 

Now let us label the eigenvalues of $I(t)$ by $\lambda_n$ and the degree of 
degeneracy of $\lambda_n$ by $l_n$. Furthermore, let $|\lambda_n,a;t\xkt$ be 
arbitrary orthonormal eigenvectors of $I(t)$ satisfying
	\xbea
	&&I(t)|\lambda_n,a;t\xkt=\lambda_n|\lambda_n,a;t\xkt\;,
	\label{ei-va-eq-I}\\
	&&\br\lambda_m,b;t|\lambda_n,a;t\kt=\delta_{mn}\delta_{ba}\;,
	\label{ortho}\\
	&&\sum_n\sum_{a=1}^{l_n}|\lambda_n,a;t\kt\br\lambda_n,a;t|=1\;,
	\label{complete}
	\xeea
where $a$ and $b$ are degeneracy labels taking their values in $\{1,2,\cdots,l_n\}$.

Clearly unlike the eigenvalues $\lambda_n$ and the corresponding degeneracy
subspaces ${\cal H}_{\lambda_n}(t)$, the eigenvectors $|\lambda_n,a;t\xkt$ are
not uniquely determined by the eigenvalue equation~(\ref{ei-va-eq-I}). They
are only determined up to unitary transformations of the degeneracy
subspaces ${\cal H}_{\lambda_n}(t)$,
	\be
	|\lambda_n,a;t\xkt\to |\lambda_n,a;t\xkt'=\sum_{b=1}^{l_n}
	|\lambda_n,b;t\xkt u_{ba}(t)\;,
	\label{trans}
	\ee
where $u_{ab}(t)$ are the entries of an arbitrary unitary $l_n\times l_n$ matrix 
$u(t)$.

The main result of Lewis and Riesenfeld \cite{le-ri} is that one can choose
a particular set of eigenvectors $|\lambda_n,a;t\xkt'$ that are solutions of 
the Schr\"odinger equation (\ref{sch-eq}). These eigenvectors are given by 
\cite{jpa-98c}:
	\bea
	|\lambda_n,a;t\kt'&:=&\sum_{b=1}^{l_n}|\lambda_n,a;t\kt u^n_{ab}(t)\;,
	~~~~{\rm where}
	\label{solu} \\	
	u^n(t)&:=&{\cal T} e^{-i\int_0^t\Delta^n(t')dt'}u^n(0)=
	{\cal T} e^{-i\int_0^t[{\cal E}^n(t')-{\cal A}^n(t')]dt'} u^n(0)\;,
	\label{u=}
	\eea
${\cal T}$ stands for the time-ordering operator and $\Delta^n(t)$, 
${\cal E}^n(t)$ and ${\cal A}^n(t)$ are Hermitian $l_n\times l_n$ matrices with 
entries
	\bea
	\Delta^n_{ab}(t)&:=&{\cal E}^n_{ab}(t)-{\cal A}^n_{ab}(t)\;,
	\label{D=}\\
	{\cal E}^n_{ab}(t)&:=&\xbr \lambda_n,a;t|H(t)|\lambda_n,b;t\xkt\;,~~~{\rm and}
	\label{E=}\\
	{\cal A}^n_{ab}(t)&:=&i\br\lambda_n,a;t|\frac{d}{dt}|\lambda_n,b;t\xkt\;,
	\label{A=}
	\eea
respectively. In other words, $u^n(t)$ is a solution of the matrix Schr\"odinger equation
	\be
	i\frac{d}{dt} u^n(t)=\Delta^n(t)u^n(t)\;. 
	\label{sch-eq-u}
	\ee

If the matrices ${\cal E}^n(t)$ and ${\cal A}^n(t)$ commute, then we can write
	\be
	u^n(t)={\cal T} e^{-i\int_0^t{\cal E}^n(t')dt'}
	{\cal T} e^{-i\int_0^t{\cal A}^n(t')dt'}u^n(0)\;.
	\label{u=1}
	\ee

For the case where the invariant $I(t)$ is periodic \cite{mkn,jpa-98c}, i.e., 
$I(T)=I(0)$ for some $T$, $|\lambda_n,a;T\xkt=|\lambda_n,a;0\xkt$ and 
$|\lambda_n,a;t\xkt'$ undergoes a cyclic evolution. The corresponding non-Abelian 
nonadiabatic cyclic geometric phase  \cite{anandan} is given by $\Gamma^n(T)$ where
$\Gamma^n(t)$ is defined to be the unique solution of the matrix Schr\"odinger 
equation
	\be
	i\frac{d}{dt} \Gamma^n(t)=-{\cal A}^n(t)\Gamma^n(t)\;,~~~~\Gamma^n(0)=1\;.
	\label{sch-eq-G}
	\ee
Alternatively,
	\be
	\Gamma^n(t):={\cal T} e^{i\int_0^t{\cal A}^n(t')dt'}\;.
	\label{G=}
	\ee
If the eigenvalue $\lambda_n$ is nondegenerate, then $l_n=1$ and we have
	\bea
	u^n(t)&=&e^{i\delta_n(t)}\Gamma^n(t)\;,~~~~{\rm where}
	\label{u-abelian}\\
	\delta_n(t)&:=&-\int_0^t{\cal E}^n(t')dt'=-\int_0^t
	\xbr \lambda_n;t'|H(t')|\lambda_n;t'\xkt dt'\;,
	\label{delta=}\\
	\Gamma^n(t)&=&e^{i\gamma_n(t)}\;,~~~~{\rm and}~~~~
	\gamma_n(t):=\int_0^t{\cal A}^n(t')dt'=\int_0^t i 
	\xbr \lambda_n;t'|\frac{d}{dt'}|\lambda_n;t'\xkt dt'\;.
	\label{G-abelian}
	\eea
In this case $\Gamma^n(T)$ is the Abelian nonadiabatic cyclic geometric phase
\cite{a-a}.

\section{An Alternative Approach to Nonadiabatic Cyclic Geometric Phase} 

In order to make the geometric character of the cyclic geometric phase more
transparent, we shall express the invariant $I(t)$ as a linear combination
of a set of linearly independent constant Hermitian operators $X_i$,
	\be
	I(t)=\sum_{i=1}^N \theta^i(t)X_i\;.
	\label{I=parameter}
	\ee
Here $N$ is a fixed nonnegative integer, the coefficients $\theta^i$ are 
real-valued functions, and $X_i$ are generators of the group $U({\cal H})$ 
of unitary transformations of the Hilbert space ${\cal H}$. If the
system has a finite-dimensional dynamical group $G$, then $X_i$ are the 
representations of the generators of $G$. In this case $N$ is just the 
dimension of $G$. However, if the Hilbert space is infinite-dimensional, 
then in principle one will need an infinite number $N$ of generators $X_i$
of $U({\cal H})$ to satisfy (\ref{I=parameter})\footnote{Note that for
each value of $t$ there is a finite number $N$ of generators $X_i$ for which 
(\ref{I=parameter}) holds. However, for an infinite-dimensional Hilbert 
space, as $t$ changes $N$ might not have an upper bound. Therefore, in order 
to satisfy (\ref{I=parameter}) one would in general need to include an 
infinite number of generators of the group $U({\cal H})$.}, and one must 
find a way to make the right-hand side of (\ref{I=parameter}) well-defined. 
We shall not be concerned with subtleties of the infinite-dimensional unitary 
group \cite{u-infinity}, and assume that $N$ is finite.

Now since the time-dependence of $I(t)$ is governed by those of the 
parameters $\theta^i$, we can consider the parameter-dependent operator
$I[\theta]$ with eigenvalues\footnote{Note that the eigenvalues of an 
invariant operator are constant \cite{le-ri,jpa-98c}.} $\lambda_n$ and 
eigenvectors $|\lambda_n,a;\theta\kt$, and write 
	\be
	I(t)=I[\theta(t)]~~~~{\rm and}~~~~|\lambda_n,a;t\kt=
	|\lambda_n,a;\theta(t)\kt\;.
	\label{lambda-parameter}
	\ee
Here $\theta$ stands for $(\theta^1,\theta^2,\cdots,\theta^N)$. The parameters
$\theta^i$ may be viewed as local coordinates of a parameter space ${\cal M}$.
As time progresses they trace a curve ${\cal C}$ in ${\cal M}$.

If the system possesses a dynamical group $G$, then we can introduce the
parameter-dependent Hamiltonian 
	\be
	H[R]=\sum_{i=1}^N R^i X_i\;,
	\label{H-parameter}
	\ee
and identify $H(t)$ with $H[R(t)]$, \cite{berry-1984}. This means that the 
parameter space ${\cal M}$ of the invariant (\ref{I=parameter}) is the same 
as the parameter space of the Hamiltonian (\ref{H-parameter}). In this case,
we can use the results of Ref.~\cite{jmp-96} to identify ${\cal M}$ with a 
submanifold of the flag manifold $G/T$ where $T$ is a maximal torus of $G$. 

Now suppose that the curve ${\cal C}$ traced by the parameters $\theta(t)$
lies in a local coordinate patch of the parameter space ${\cal M}$. In
this case, we can introduce the nonadiabatic analogue of the non-Abelian Berry 
connection one-form \cite{wi-ze,jpa-93},
	\be
	A^n[\theta]=\sum_{i=1}^N A^n_i[\theta]d\theta^i\;.
	\label{an-comp}
	\ee
The matrix elements of $A^n[\theta]$ and its components $A^n_i[\theta]$
are given by
	\bea
	A^n_{ab}[\theta]&:=&i\br\lambda_n,a;\theta|d|\lambda_n,b;\theta\kt\;,
	~~~{\rm and}
	\label{nonadiabatic-connetion}\\
	\left(A_{i}^n[\theta]\right)_{ab}&:=&i\br\lambda_n,a;\theta|
	\frac{\partial}{\partial\theta^i}|\lambda_n,b;\theta\kt\;,
	\label{nonadiabatic-connetion-2}
	\eea
respectively. In Eq.~(\ref{nonadiabatic-connetion}) $d$ stands for the exterior 
derivative with respect to $\theta^i$. In view of Eqs.~(\ref{A=}), 
(\ref{lambda-parameter}), (\ref{an-comp}), (\ref{nonadiabatic-connetion}), and 
(\ref{G=}), we have
	\bea
	{\cal A}^n(t) dt&=&\sum_{i=1}^N A^n_i[\theta(t)]\frac{d\theta^i(t)}{dt}dt=
	A^n[\theta(t)]\;,~~~~{\rm and}
	\label{A=A}\\
	\Gamma^n(t)&=&{\cal P}e^{i\int_{\theta(0)}^{\theta(t)}A^n[\theta]}\;,
	\label{G-open}
	\eea
where ${\cal P}$ stands for the path-ordering operator. In particular, for a 
periodic invariant, the curve ${\cal C}$ traced by $\theta(t)$ is closed, and
the non-Abelian cyclic geometric phase takes the form
	\be
	\Gamma^n(T)={\cal P}e^{i\oint_{\cal C} A^n[\theta]}\;.
	\label{G=2}
	\ee
As pointed out in Ref.~\cite{jpa-98c}, this expression agrees with Anandan's 
definition of non-Abelian cyclic geometric phase \cite{anandan}.

If ${\cal C}$ does not lie in a single coordinate patch of ${\cal M}$, one must 
evaluate the path-ordered integrals in (\ref{G-open}) and (\ref{G=2}) along
the segments of ${\cal C}$ belonging to different coordinate patches and
multiply the resulting unitary matrices in the order in which the curve 
${\cal C}$ is traversed in time.

\section{Evolution Operator and Its Gauge Invariance}

In general, we can write any solution of the Schr\"odinger equation (\ref{sch-eq})
as a linear combination of $|\lambda_n,b;t\xkt'$, i.e.,
	\be
	|\psi(t)\kt=\sum_n\sum_{a=1}^{l_n}\tilde C^n_a|\lambda_n,a;t\xkt'
	=\sum_n\sum_{a,b=1}^{l_n} C^n_a u^n_{ba}(t)|\lambda_n,b;t\xkt\;,
	\label{psi=}
	\ee
where 
	\be
	\tilde C^n_a:=\sum_{b=1}^{l_n}u^{n\dagger}_{ba}(0)
	\br\lambda_n,b;t|\psi(0)\kt\;,~~~{\rm and}~~~
	C^n_a:=\br\lambda_n,a;t|\psi(0)\kt\;.
	\label{cc}
	\ee
In view of Eq.~(\ref{psi=}), the evolution operator is given by
	\be
	U(t)=\sum_n\sum_{a,b=1}^{l_n}u^n_{ab}(t)|\lambda_n,a;t\kt\br\lambda_n,b;0|\;.
	\label{evo-op}
	\ee

Now let us recall that the eigenvectors $|\lambda_n,a;\theta\kt$ are not uniquely
determined by the eigenvalue equation
	\be
	I[\theta]|\lambda_n,a;\theta\kt=\lambda_n|\lambda_n,a;\theta\kt\;.
	\label{eg-va-eq-I-theta}
	\ee
They are subject to arbitrary gauge transformations
	\be
	|\lambda_n,a;\theta\kt\to|\lambda_n,a;\theta\kt\tilde{~}:=\sum_{b=1}^{l_n}
	|\lambda_n,b;t\xkt v^n_{ba}[\theta]\;,
	\label{gauge-trans}
	\ee
where $v^n_{ab}[\theta]$ are entries of an $l_n\times l_n$ unitary matrix $v^n[\theta]$.
In fact, if we denote the local coordinate patch corresponding to the coordinates 
$\theta^i$ by ${\cal O}$, $v^n$ may be viewed as a smooth function mapping ${\cal O}$ 
to the unitary group $U(l_n)$. 

Using Eq.~(\ref{nonadiabatic-connetion}), we can easily derive the gauge transformation
law for $A^n[\theta]$, namely
	\be
	A^n[\theta]\to\tilde A^n[\theta]=v^n[\theta]^\dagger A^n[\theta] v^n[\theta]
	+iv^n[\theta]^\dagger d v^n[\theta]\;.
	\label{gauge-trans-A}
	\ee
Furthermore, in view of Eqs.~(\ref{A=A}), (\ref{sch-eq-G}), and (\ref{gauge-trans-A}),
we have the following transformation rules for $\Gamma^n(t)$ and $u^n(t)$.
	\bea
	\Gamma^n(t)&\to&\tilde\Gamma^n(t)=v^n[\theta(t)]^\dagger \Gamma^n(t) 
	v^n[\theta(0)]\;,~~~{\and}
	\label{gauge-trans-G}\\
	u^n(t)&\to&\tilde u^n(t)=v^n[\theta(t)]^\dagger u^n(t) 
	v^n[\theta(0)]\;.
	\label{gauge-trans-u}
	\eea
A simple consequence of Eqs.~(\ref{gauge-trans}) and (\ref{gauge-trans-u})
is that the evolution operator (\ref{evo-op}) is invariant under the gauge 
transformations.

For a periodic invariant $I(t)$, with $\theta(T)=\theta(0)$, $\tilde\Gamma^n(T)$
is related to $\Gamma^n(T)$ by a similarity transformation
	\be
	\tilde\Gamma^n(T)=v^n[\theta(0)]^\dagger \Gamma^n(T) v^n[\theta(0)]\;.
	\label{trans-G-T}
	\ee
In other words, under a gauge transformation (\ref{gauge-trans}) $\Gamma^n(T)$ 
transforms covariantly. Therefore, its eigenvalues and in particular its trace
are gauge-invariant. These are essentially the physically observable
quantities associated with the non-Abelian cyclic geometric phase.

\section{Noncyclic Geometric Phase}

The main reason for Berry's consideration of a periodic Hamiltonian 
\cite{berry-1984} and Aharonov and Anandan's consideration of cyclic evolutions
\cite{a-a} is the fact that for a cyclic state with period $T$-- in our 
approach a $T$-periodic dynamical invariant -- the unitary matrix $\Gamma^n(T)$
transforms convariantly under a gauge transformation (\ref{gauge-trans}). This 
property of $\Gamma^n(T)$ guarantees that its eigenvalues and its trace are 
gauge-invariant. This in turn raises the issue of exploring their physical 
consequences. There is, however, another way of constructing gauge-invariant
quantities using $\Gamma^n(t)$.

In view of Eqs.~(\ref{gauge-trans-G}) and (\ref{gauge-trans-u}), $\Gamma^n(t)$
and $u^n(t)$ have the same gauge transformation properties. This means that if
we replace $u^n(t)$ in the expression (\ref{evo-op}) for the evolution operator
by $\Gamma^n(t)$, then we shall still obtain a gauge-invariant operator, namely
	\be
	V(t):=\sum_n\sum_{a,b=1}^{l_n}\Gamma^n_{ab}(t)|\lambda_n,a;\theta(t)\kt
	\br\lambda_n,b;\theta(0)|\;.
	\label{V}
	\ee
In fact, because the gauge transformations (\ref{gauge-trans}) are the
unitary transformations of the degeneracy subspaces ${\cal H}_{\lambda_n}[\theta(t)]$, 
the restriction (or projection) of $V(t)$ onto the degeneracy subspaces, i.e.,
	\be
	V^n(t):=\sum_{a,b=1}^{l_n}\Gamma^n_{ab}(t)|\lambda_n,a;\theta(t)\kt
	\br\lambda_n,b;\theta(0)|\;,
	\label{Vn}
	\ee
will also be gauge-invariant. By construction, the operators $V^n(t)$ are uniquely 
determined by the curve ${\cal C}$ and its end points. In particular, they are 
independent of the duration of the evolution. Therefore, they are also geometric
quantities. 

Next let us note that the gauge-invariance and geometric character of $V^n(t)$ 
will not be affected, if we exchange the positions of $|\lambda_n,a;\theta(t)\kt$ and
$\br\lambda_n,b;\theta(0)|$ in Eq.~(\ref{Vn}). In this way we obtain a set of
gauge invariant scalars:
	\be
	\Pi^n(t):=\sum_{a,b=1}^{l_n}\Gamma_{ab}^n(t)\br\lambda_n,b;\theta(0)|
	\lambda_n,a;\theta(t)\kt=\sum_{a,b=1}^{l_n}w^n_{ba}(t)\Gamma_{ab}^n(t)
	={\rm trace}[w^n(t)\Gamma^n(t)]\;,
	\label{Pi}
	\ee
where we have introduced the $l_n\times l_n$ matrices $w^n(t)$ with entries
	\be
	w^n_{ba}(t):=\br\lambda_n,b;\theta(0)|\lambda_n,a;\theta(t)\kt\;.
	\label{w=}
	\ee
By definition $w^n(t)$ only depend on the end points of the curve ${\cal C}$.
$\Gamma^n(t)$ are also uniquely determined by ${\cal C}$. Therefore, as
expected $\Pi^n(t)$ are geometric quantities.

Now let us consider a $T$-periodic invariant (a cyclic evolution), for which
$|\lambda_n,a;\theta(T)\kt=|\lambda_n,a;\theta(0)\kt$. In this case,  $w^n(T)=
w^n(0)$ is just the identity matrix, and  Eq.~(\ref{Pi}) reduces to
	\be
	\Pi^n(T)={\rm trace}[\Gamma^n(T)]\;.
	\label{trace}
	\ee
Therefore, for a cyclic evolution $\Pi^n(T)$ yields the trace of the non-Abelian 
cyclic geometric phase \cite{wi-ze,zee}.

On the other hand, since $\Pi^n(t)$ is gauge-invariant, we can compute it in
a basis $\{|\lambda_n,a;\theta(t)\kt_\star\}$ in which $\Gamma^n(t)$ is diagonal. 
If we denote the eigenvalues of $\Gamma^n(t)$ by $e^{i\gamma_n^a(t)}$, then 
we  have
	\be
	\Pi^n(t)=\sum_{a=1}^{l_n}e^{i\gamma_n^a(t)}\,
	{~}_\star\br\lambda_n,a;\theta(t)|\lambda_n,a;\theta(0)\kt_\star\;.
	\label{diag}
	\ee
As seen from Eq.~(\ref{diag}), in general $\Pi^n(t)$ is a nonunimodular complex 
number. 

In view of the above analysis, the matrix $\check{\Gamma}^n({\cal C})$
defined by
	\be
	\check{\Gamma}^n({\cal C})=\check{\Gamma}^n(t)
	:=w^n(t)\Gamma^n(t)\;,
	\label{noncyclic-ge-ph}
	\ee
is a gauge-covariant geometric quantity. We shall therefore identify it as the 
{\em non-Abelian noncyclic geometric phase factor}. Clearly, the eigenvalues 
of $\check{\Gamma}^n({\cal C})$ and its trace namely $\Pi^n(t)$
are gauge-invariant quantities which can, in principle, be observed
experimentally. The non-Abelian noncyclic geometric phase factor
$\check{\Gamma}^n({\cal C})$ is therefore as physically significant as its
cyclic counterpart (\ref{G=2}). 

As mentioned above, for a cyclic evolution where the invariant $I(t)$ is 
$T$-periodic, $|\lambda_n,a;\theta(T)\kt=|\lambda_n,a;\theta(0)\kt$ and
$w^n(T)$ is the identity operator. In this case, $\check{\Gamma}^n(T)$ is
identical with the cyclic geometric phase factor $\Gamma^n(T)$ given
by Eq.~(\ref{G=2}). 

\section{Abelian Noncyclic Geometric Phase}

For a nondegenerate eigenvalue $\lambda_n$ of the invariant $I(t)$, we
have
	\be
	\check{\Gamma}^n(t) =w^n(t)\Gamma^n(t)=\Pi^n(t)=
	\br\lambda_n;\theta(0)|\lambda_n;\theta(t)\kt
	e^{i\gamma_n(t)}\;,
	\label{Pi-a}
	\ee
where $\gamma_n(t)$ is the phase angle given by (\ref{G-abelian}). In particular,
	\be
	|\check{\Gamma}^n(t)|=|w^n(t)|=
	|\br\lambda_n;\theta(0)|\lambda_n;\theta(t)\kt|
	\label{mag}
	\ee
depends only on the end points of the curve ${\cal C}$. If 
$|\check{\Gamma}^n(t)|\neq 0$, then we can consider the phase of 
$\check{\Gamma}^n(t) $ which is given by
	\bea
	e^{i\check{\gamma}_n(t)}&:=&\frac{\check{\Gamma}^n(t)}{
	|\check{\Gamma}^n(t)|}=e^{i[\eta_n(t)+\gamma_n(t)]}
	\;,~~~{\rm where}~~~
	\label{pi}\\
	e^{i\eta_n(t)}&:=&\frac{w^n(t)}{|w^n(t)|}=
	\frac{\br\lambda_n;\theta(0)|\lambda_n;\theta(t)\kt}{
	|\br\lambda_n;\theta(0)|\lambda_n;\theta(t)\kt|}=
	\left[\frac{\br\lambda_n;\theta(0)|\lambda_n;\theta(t)\kt }{
	\br\lambda_n;\theta(t)|\lambda_n;\theta(0)\kt}\right]^{1/2}\;.
	\label{phase}
	\eea
The phase angle $\check{\gamma}_n(t)$ is a {\em real noncyclic geometric 
phase angle}. It consists of two pieces: $\gamma_n(t)$ that depends on the curve 
${\cal C}$, and $\eta_n(t)$ that depends on the end points of ${\cal C}$.

The phase factor (\ref{pi}) coincides with the `real noncyclic geometric phase' 
introduced by Aitchison and Wanelik \cite{ai-wa}. As discussed by Aitchison and 
Wanelik it is equivalent to the noncyclic geometric phase of Samuel and Bhandari
\cite{sa-bh} and Mukunda and Simon \cite{mu-si}.

\section{Adiabatic Approximation and the Noncyclic Geometric Phase in
the Adiabatic Limit}

Let $H[R]$ be a parameter-dependent Hamiltonian with a discrete spectrum.
We shall denote the eigenvalues of $H[R]$ by $E_n[R]$, their degree of
degeneracy by $\cun$, and the corresponding degeneracy subspaces by
${\cal H}_n[R]$. Let $|n,a;R\kt$ form a complete set of orthonormal 
eigenvectors of $H[R]$. They satisfy,
	\be
	H[R]|n,a;R\kt=E_n[R]|n,a;R\kt\;,
	\label{eg-va-H}
	\ee
where $a$ is a degeneracy label taking its value in $\{1,2,\cdots,\cun\}$.

Now consider the time-dependent Hamiltonian $H(t):=H[R(t)]$, where the
parameters $R(t)=(R^1(t),R^2(t),\cdots,R^d(t))$ trace a curve $C$ in the 
parameter space $M$ of the Hamiltonian. We shall denote the duration of 
the evolution of the system by $\tau$ and suppose that during the evolution 
no level-crossings occur. Furthermore, let $I(t)$ be a dynamical invariant 
satisfying (\ref{dyn-inv}), and suppose that it depends on a set of parameters
$\theta=(\theta^1,\theta^2,\cdots,\theta^N)$, i.e., $I(t)=I[\theta(t)]$ where
$\theta(t)$ traces a curve ${\cal C}$ in the parameter space ${\cal M}$ of 
the invariant. Since $I(t)$ yields the solution of the Schr\"odinger equation,
the dynamics of the system can be encoded in the definition of a function
	\be
	F:M\to{\cal M},~~~{\rm defined~by}~~~F(R):=\theta.
	\label{F}
	\ee
In particular, $F$ maps the curve $C$ onto the curve ${\cal C}$, and 
$I(t)=I[\theta(t)]=I[F(R(t))]$. Note, however, that there does not exist
a universal function $F$ describing all possible dynamical processes.
In other words, the definition of $F$ also depends on the choice of the
Hamiltonian or alternatively the curve $C$. In this sense, it is more appropriate
to define the function
	\be
	{\cal F}:{\cal P}_M\to{\cal P}_{\cal M}~~~{\rm by}~~~
	{\cal F}(C)={\cal C}\;.
	\label{cal-F}
	\ee
where ${\cal P}_M$ and ${\cal P}_{\cal M}$ are the space of paths in $M$
and ${\cal M}$ respectively.	The latter are infinite-dimensional spaces. Therefore,
it is more convenient to use the function $F$ with the provision of its nonuniversal
character.

Next we define a normalized time variable by $s:=t/\tau$, and assume that 
for sufficiently large values of $\tau$ we can expand $I(t)$ and $H(t)$ in 
powers of $\tau^{-1}$, i.e.,
	\be
	I(t)=I(\tau s)=I_0(s)+\sum_{\ell=1}^\infty \tau^{-\ell}I_\ell(s)~~~
	{\rm and}~~~
	H(t)=H(\tau s)=H_0(s)+\sum_{\ell=1}^\infty \tau^{-\ell}H_\ell(s)\;,
	\label{I-H}
	\ee
where $I_\ell$ and $H_\ell$ are Hermitian operators and $I_0\neq 0\neq H_0$.
If we substitute $t=\tau s$ and (\ref{I-H}) in (\ref{dyn-inv}) and take the limit
$\tau\to\infty$, we find $\left[I_0(s),H_0(s)\right]=0$. This means that in this
limit, where $I(t)\to I_0(s)$ and $H(t)\to H_0(s)$, the eigenvectors of $I(t)$ and
$H(t)$ coincide. Using the notation of the preceding sections, we have
	\be
	\lim_{\tau\to\infty}|\lambda_n,a;t\kt=|n,a;t\kt \;.
	\label{l=n}
	\ee
Because $|\lambda_n,a;t\kt=|\lambda_n,a;\theta(t)\kt$, $|n,a;t\kt=|n,a;R(t)\kt$, and
(\ref{l=n}) is independent of the form of the curve $C$,
	\be
	\lim_{\tau\to\infty}\left.|\lambda_n,a;\theta\kt\right|_{\cal C}=
	\left.|n,a;R\kt\right|_{C}\;.
	\label{lim-cc}
	\ee
In particular, this implies that for sufficiently large values of $\tau$, we can
choose an invariant whose parameter space is the same as that of the 
Hamiltonian, ${\cal M}=M$. Therefore, $C$ and ${\cal C}$ belong to the
same parameter space $M$. In this case, we can also express (\ref{l=n})
and (\ref{lim-cc}) by
	\be
	\lim_{\tau\to\infty}F={\rm id}_M\;.
	\label{lim-F}
	\ee
where ${\rm id}_M$ is the identity function on $M$.

Now let us use Eqs.~(\ref{evo-op}) and (\ref{l=n}) to compute the evolution 
operator in the limit $\tau\to\infty$. This leads to
	\bea
	&&\lim_{\tau\to\infty}U(t)=U^{(0)}(t)\;,~~~{\rm where}~~~
	\label{lim-u}\\
	&&U^{(0)}(t):=\sum_n\sum_{a,b=1}^{\cun} u^n_{\circ ab}(t)|n,a;t\kt\br n,b;0|\;,
	\label{u0}\\
	&&u_\circ^n(t)=e^{i\delta_{\circ n}(t)}\Gamma^n_\circ (t)\;,
	\label{un-0}\\
	&&\delta_{\circ n}(t):=-\int_0^t E_n(t')dt'\;,
	\label{delta-0}\\	
	&&\Gamma^n_\circ(t):={\cal P}e^{i\int_{R(0)}^{R(t)}A^n_\circ[R]}\;,
	~~~{\rm and}
	\label{G-0}\\
	&&\left( A^n_\circ[R]\right)_{ab}:=i\br n,a;R|d|n,b;R\kt\;.
	\label{A-0}
	\eea
Here the subscript $\circ$ is inserted to mean that the corresponding quantities
are obtained in the adiabatic limit ($\tau\to\infty$) from the ones without a
subscript $\circ$. The matrix of one-forms $A^n_\circ[R]$ is the non-Abelian
Berry connection one-form \cite{wi-ze}.

In practice, $\tau$ is a finite number and the limit $\tau\to\infty$ is interpreted
by the condition that $\tau$ must be much larger than the time  (inverse of
energy) scale of the quantum system. If this happens to be the case the above
results may be used. It is not difficult  to check that an operator $I(t)$ with the
same eigenvectors as the Hamiltonian satisfies Eq.~(\ref{dyn-inv}) only if the
eigenvectors of the Hamiltonian are constant. This is often not the case. Therefore,
in an eigenbasis $\{|n,a;t\kt\}$ of the Hamiltonian, the invariant $I(t)$ is not
diagonal. However, if the above adiabaticity condition is fulfilled, i.e., $\tau$
is much larger than the time scale of the problem, then the off-diagonal matrix
elements of $I(t)$ are much smaller than its diagonal matrix elements. The
approximation scheme in which one neglects the off-diagonal matrix elements
of $I(t)$ is called the {\em adiabatic approximation}, \cite{bo-fo,kato,pra-97a}. 
In this approximation, we have
	\be
	|\lambda_n,a;t\kt\approx|n,a;t\kt\;,~~~U(t)\approx U^{(0)}(t)
	\;,~~~{\rm and}~~~F\approx{\rm id}_M\;.
	\label{adi-app}
	\ee
This is a valid approximation scheme if and only if 
	\be
	{\cal A}^{mn}_{ba}(t):=i\br m,b;R|\frac{d}{dt}|n,a;R\kt
	=\frac{i\br m,b;R|\frac{dH(t)}{dt}|n,a;R\kt}{E_n(t)-E_m(t)}\approx 0
	~~~{\rm for}~~~m\neq n\;.
	\label{condi}
	\ee
Here $m$ and $n$ are arbitrary labels (satsifying $m\neq n$) and $a$ and $b$ are 
arbitrary degeneracy labels associated with the eigenvalues $E_n(t)$ and $E_m(t)$, 
respectively. The second equation in (\ref{condi}) is obtained by differentiating both 
sides of Eq.~(\ref{eg-va-H}) with respect to time and taking the inner product of both 
sides of the resulting equation with $|m,b;t\kt$. The meaning of `$\approx 0$' in 
(\ref{condi}) is that the left-hand side of (\ref{condi}) which has the dimension of 
frequency must be much smaller than the frequency (energy) scale of the system.

In view of (\ref{adi-app}), we have the following expression for the adiabatic
non-Abelian noncyclic geometric phase (\ref{noncyclic-ge-ph}),
	\be
	\check{\Gamma}^n_\circ(t)=w_\circ^n(t)\Gamma_\circ^n(t)\;,
	\label{Pi-0}
	\ee
where $w_\circ^n(t)$ is an $\cun\times\cun$ matrix with entries
	\be
	w^n_{\circ ab}(t):=\br n,a;R(0)|n,b;R(t)\kt\;.
	\label{w-0=}
	\ee

If $E_n[R]$ is nondegenerate, $\cun=1$ and 
	\be
	\check{\Gamma}^n_\circ(t)=w_\circ^n(t) e^{i\gamma_{\circ n}(t)}\;,~~~
	{\rm where}~~~
	\gamma_{\circ n}(t):=\int_{R(0)}^{R(t)} A^n_\circ[R]=
	\int_{R(0)}^{R(t)} i\br n;R|d|n;R\kt\;.
	\label{abelian-Pi-0}
	\ee
The phase of $\check{\Gamma}^n_\circ(t)$, namely 
	\be
	e^{i\check{\gamma}_{\circ n}(t)}=\left[\frac{\br n;R(0)|n;R(t)\kt}{\br n;R(t)
	|n;R(0)\kt}\right]^{1/2}e^{i\gamma_{\circ n}(t)}\;,
	\label{phase-0}
	\ee
is precisely the Abelian adiabatic noncyclic geometric phase studied by
de~Polavieja and Sj\"oqvist~\cite{po-sj}.

\section{Application: Spin 1 Quadrupole in a Precessing Magnetic Field}

The simplest possible quantum system that allows for the occurrence of a non-Abelian
geometric phase is a system with a three-dimensional Hilbert space and a dynamical
invariant $I(t)$ which has a nondegenerate and a degenerate eigenvalue \cite{jpa-98c}.
If the system undergoes an adiabatic evolution, then the role of the invariant is 
essentially played by the Hamiltonian. In particular, the Hamiltonian must have a 
nondegenerate and a degenerate eigenvalue. The moduli space of all such Hamiltonians
(for a three-dimensional Hilbert space) has the manifold structure of the projective 
space $\xC P^2$, \cite{wi-ze}. A thorough treatment of the problem of the adiabatic 
non-Abelian cyclic geometric phase for such a system is presented in Ref.~\cite{jpa-97}. 
In this section, we shall use the results of  Ref.~\cite{jpa-97} to investigate the 
adiabatic non-Abelian noncyclic geometric phase $\check{\Gamma}^n_\circ(t)$ for 
a spin 1 quadrupole interacting with a precessing magnetic or electric field. The 
problem of the non-Abelian geometric phase for a spin 3/2 quadrupole has been 
considered by Zee \cite{zee}, Mead \cite{mead} and Avron, Sadun, Segert, and 
Simon \cite{av-sa-se-si-prl,av-sa-se-si-cmp}. For a bosonic system such as the spin 1 
quadrupole considered here, one can show that Berry's connection one-form is a 
pure gauge and Berry's cyclic geometric phase angle vanishes \cite{av-sa-se-si-cmp}.
This result does not however generalize to the non-Abelian geometric phase,
\cite{jpa-97}.

Consider the quadrupole Stark Hamiltonian $H=\lambda( J\cdot R)^2$,
where $\lambda$ is a real coupling constant, $J=(J_1,J_2,J_3)$ is the
angular momentum operator, and $R=(R^1,R^2,R^3)$ is a 3-vector representing
the magnetic (or the electric) field. For a spin 1 particle, this Hamiltonian has the form
	\be
	H[R]=\frac{\lambda\rho}{2}\left(\begin{array}{ccc}
	1+2\zeta^2 & \sqrt{2}~ \zeta e^{-i\varphi} & e^{-2i\varphi} \\
	 \sqrt{2}~\zeta e^{i\varphi} & 2 & -\sqrt{2}~\zeta e^{-i\varphi}\\
	e^{-2i\varphi} & -\sqrt{2}~\zeta e^{i\varphi}&1+2\zeta^2
	\end{array}\right)\;,
	\label{qua-H}
	\ee
where $(\rho,\varphi,z)$ are the cyclindrical coordinates in the $R$-space, i.e.,
	\[\rho:=\sqrt{(R^1)^2+(R^2)^2}\;,~~~\varphi:=\tan^{-1}(R^2/R^1)\;,
	~~~z:=R^3\;,\]
$\zeta:=z/\rho$, and we have used the standard spin 1 representations of $J_i$. 

In view of the general results of Ref.~\cite{jpa-97}, the eigenvalues of the Hamiltonian
(\ref{qua-H}) are given by
	\be
	E_1[R]=\lambda\rho^2\zeta^2\;,~~{\rm and}~~E_2[R]=
	\lambda\rho^2(1+\zeta^2)\;,
	\label{e1-e2}
	\ee
where $R=(\rho,\varphi,\zeta)$. As seen from (\ref{e1-e2}), if $\rho\neq 0$, then the 
Hamiltonian has two distinct eigenvalues. In this case, $E_1[R]$ is nondegenerate and 
$E_2[R]$ is doubly degenerate. A set of orthonormal eigenvectors of $H[R]$ is given 
by \cite{jpa-97}
	\bea
	|1;R\kt&:=&N_1^{-1}\left(\begin{array}{c}
	e^{-i\varphi}\\
	\sqrt{2}~\zeta\\	
	e^{i\varphi}\end{array}\right)\;,\nn\\
	|2,1;R\kt&:=&N_2^{-1}\left(\begin{array}{c}
	-\sqrt{2}~\zeta e^{-i\varphi}\\
	1\\	
	0\end{array}\right)\;,~~~{\rm and}~~~
	|2,2;R\kt:=(N_1N_2)^{-1}\left(\begin{array}{c}
	e^{-i\varphi}\\
	\sqrt{2}~\zeta\\	
	(1+2\zeta^2)e^{i\varphi}\end{array}\right)\;,
	\label{qua-ei-vec}
	\eea
where $N_1:=\sqrt{2(1+\zeta^2)}$ and $N_2:=\sqrt{1+2\zeta^2}$. Note that these
formulas are valid for $\rho\neq 0$, i.e., everywhere except the $R^3$-axis.

Again using the general results of Ref.~\cite{jpa-97} or by direct calculation, we can 
compute Berry's connection one-forms $A_\circ^n$. Doing the necessary algebra, we 
find
	\be
	A_\circ^1[R]=d\varphi\;,~~~{\rm and}~~~
	A_\circ^2[R]=\left(\begin{array}{cc}
	\mu d\varphi & \nu e^{i\varphi}d\varphi \\
	 \nu e^{-i\varphi}d\varphi &\sigma d\varphi\end{array}\right)\;,
	\label{A1-A2}
	\ee
where
	\bea
	\mu&:=&\frac{2\zeta^2}{1+2\zeta^2}=\frac{2 c^2}{1+c^2}\;,
	\label{mu}\\
	\nu&:=&-\frac{\zeta}{(1+2\zeta^2)\sqrt{1+\zeta^2}}=-\frac{c(1-c^2)}{1+c^2}\;,
	\label{nu}\\
	\sigma&:=&-\frac{1+2(1+2\zeta^2)^2}{2(1+2\zeta^2)(1+\zeta^2)}=
	-\frac{1+2(1+c^2)^2}{2(1+c^2)}\;,
	\label{sigma}
	\eea
and $c:=\zeta/(1+\zeta^2)=z/\sqrt{\rho^2+\zeta^2}=R^3/\sqrt{(R^1)^2+(R^2)^2+
(R^3)^2}$. Note that in the spherical coordinates $(r,\theta,\varphi)$, we have
	\be
	 c=\cos\theta\;.
	\label{c=}
	\ee

In view of Eqs.~(\ref{A1-A2}), $A_\circ^1[R]$ is a pure gauge. This was to be 
expected, for the system is bosonic \cite{av-sa-se-si-cmp}. The connection one-form
$A_\circ^2[R]$ is not a pure gauge. In fact, unlike the spin 3/2 systems considered in
the literature \cite{zee} even for the case of a precessing field where $\theta$ is
kept fixed and $\varphi$ varies, the adiabatic geometric phase associated with $E_2$ is 
non-Abelian. 

In the remainder of this section, we shall compute the adiabatic non-Abelian noncyclic 
geometric phase associated with the degenerate eigenvalue $E_2$ for a precessing field with
	\be
	\theta={\rm constant}\;,~~~~\varphi=\varphi_0+\omega t\;,~~~{\rm and}~~~
	\varphi_0,\omega={\rm constant}\;.
	\label{precessing}
	\ee

First, let us consider the matrix $\Gamma^2_\circ$. We can write Eq.~(\ref{G-0})
as the matrix Schr\"odinger equation
	\be
	i\frac{d}{d\varphi}\Gamma^2_\circ(\varphi)=
	h(\varphi)\Gamma^2_\circ(\varphi)\,,
	\label{dG=-AG}
	\ee
where 
	\be
	h(\varphi)d\varphi:=-A^2_\circ[R]\;.
	\label{little-h}
	\ee
Clearly, $h(\varphi)$ belongs to the Lie algebra of the unitary group $U(2)$.
In particular, it can be written in the form
	\be
	h(\varphi)=\frac{1}{2}\sum_{\ell=0}^3 r^\ell \sigma_\ell\;,
	\label{h-u2}
	\ee
where $\sigma_0$ stands for the unit $2\times 2$ matrix, $\sigma_\ell$
with $\ell\in\{1,2,3\}$ are Pauli matrices, and
	\bea
	&&r^0:=-(\mu+\sigma)\;,~~~~r^1:=-\nu\cos\varphi\;,\nn\\
	&&r^2:=\nu\sin\varphi\;,~~~{\rm and}~~~r^3:=\sigma-\mu\;.\nn
	\eea
Substituting these equations in (\ref{h-u2}), we obtain
	\bea
	h(\varphi)&=&\frac{1}{2}[-(\mu+\sigma)\sigma_0-\nu(\cos\varphi\;\sigma_1-
	\sin\varphi\;\sigma_2)+(\sigma-\mu)\sigma_3]\;,\nn\\
	&=&\frac{1}{2}e^{i\varphi\,\sigma_3/2} [-(\mu+\sigma)\sigma_0-\nu\;
	\sigma_1+(\sigma-\mu)\sigma_3]e^{-i\varphi\,\sigma_3/2}\;,
	\label{h=ehe}
	\eea
where we have used the identity
	\be
	e^{-i\varphi\sigma_i/2}\sigma_je^{i\varphi\sigma_i/2}=
	\cos\varphi\;\sigma_j+\sin\varphi\,\sum_{k=1}^3\epsilon_{ijk}\sigma_k\;,
	~~~{\rm for}~~~i\neq j\;.
	\label{identity}
	\ee
In (\ref{identity}), $\epsilon_{ijk}$ stands for the totally antisymmetric Levi Civita 
symbol with $\epsilon_{123}=1$.
	
In view of Eq.~(\ref{h=ehe}), $h(\varphi)$ is the Hamiltonian of a spin 1/2 magnetic
dipole in a precessing magnetic field. Therefore, we can perform a unitary 
transformation of the Hilbert space \cite{ra-ra-sc,jmp-97b} to map it to a constant
Hamiltonian. Under a $\varphi$-dependent unitary transformation $ {\cal U}(\varphi) $,
$h(\varphi)$ and $\Gamma^2_\circ(\varphi) $ transform according to \cite{jmp-97b} 
	\bea
	h(\varphi)&\to&h'(\varphi)={\cal U}(\varphi)h(\varphi){\cal U}(\varphi)^\dagger
	-i{\cal U}(\varphi)\frac{d}{d\varphi}{\cal U}(\varphi)^\dagger\;,
	\label{h-trans}\\
	\Gamma^2_\circ(\varphi)&\to&\Gamma^{'2}_\circ(\varphi)=
	{\cal U}(\varphi)\Gamma^{2}_\circ(\varphi){\cal U}(\varphi_0)^\dagger\;.
	\label{u-trans}
	\eea
Setting ${\cal U}(\varphi)=e^{-i\varphi\sigma_3/2} $ in (\ref{h-trans}) and
using (\ref{h=ehe}), we find
	\be
	h'=\frac{1}{2}[-(\mu+\sigma)\sigma_0-\nu\sigma_1+(1-\mu+\sigma)
	\sigma_3]\;.
	\label{h'}
	\ee

For a precessing field, $\theta$ and consequently $c$, $\mu$, $\nu$, and $\sigma$
are constant parameters. Therefore, $h'$ is constant, and we have
	\be
	\Gamma^{'2}_\circ(\varphi)=e^{-ih'(\varphi-\varphi_0)}\;.
	\label{G'}
	\ee
Substituting this equation in (\ref{u-trans}), we find
	\be
	\Gamma^2_\circ(\varphi)=e^{i\varphi\sigma_3/2}e^{-ih'(\varphi-\varphi_0)}
	e^{-i\varphi_0\sigma_3/2}\;.
	\label{G2=}
	\ee
In view of Eqs.~(\ref{h'}) and (\ref{G2=}), the matrix elements of  
$\Gamma^2_\circ(\varphi)$ are given by
	\bea
	\Gamma^2_{\circ 11}&=& e^{i(\mu+\sigma+1)(\varphi-\varphi_0)/2}
	\left\{\cos[(\varphi-\varphi_0)\Delta/2] +i\left(\frac{\mu-\sigma-1}{\Delta}\right)
	\sin[(\varphi-\varphi_0)\Delta/2] \right\}\;,
	\label{G2-11}\\
	\Gamma^2_{\circ 12}&=&i \left(\frac{\nu}{\Delta}\right)
	e^{i(\mu+\sigma)(\varphi-\varphi_0)/2}
	e^{i(\varphi+\varphi_0)/2}\sin[(\varphi-\varphi_0)\Delta/2] \;,
	\label{G2-12}\\
	\Gamma^2_{\circ 21}&=&i \left(\frac{\nu}{\Delta}\right)
	e^{i(\mu+\sigma)(\varphi-\varphi_0)/2}
	e^{-i(\varphi+\varphi_0)/2}\sin[(\varphi-\varphi_0)\Delta/2] \;,
	\label{G2-21}\\
	\Gamma^2_{\circ 22}&=& e^{i(\mu+\sigma-1)(\varphi-\varphi_0)/2}
	\left\{\cos[(\varphi-\varphi_0)\Delta/2] -i\left(\frac{\mu-\sigma-1}{\Delta}\right)
	\sin[(\varphi-\varphi_0)\Delta/2] \right\}\;,
	\label{G2-22}
	\eea
where
	\be
	\Delta:=\sqrt{(1+\sigma-\mu)^2+\nu^2}=\frac{\sqrt{1+4c^2(
	4+8c^2+7c^4+c^6)}}{2(1+c^2)}\;.
	\label{Delta=sqrt}
	\ee

Next we compute the entries of the matrix $w^2_\circ(t)$ of Eq.~(\ref{w-0=}).
Using Eqs.~(\ref{qua-ei-vec}) and doing the necessary algebra, we have
	\bea
	w^2_{\circ 11}&=&1-\left(\frac{2c^2}{1+c^2}\right) 
	(1-e^{-i(\varphi-\varphi_0)})=1-\mu(1-e^{-i(\varphi-\varphi_0)}) \;,
	\label{w211}\\
	w^2_{\circ 12}&=&w^2_{\circ 21}=\left[\frac{c(1-c^2)}{1+c^2}\right]
	 (1-e^{-i(\varphi-\varphi_0)})=-\nu(1-e^{-i(\varphi-\varphi_0)}) \,,
	~~{\rm and}
	\label{w212} \\
	w^2_{\circ 22}&=&1-\left(\frac{1+c^4}{1+c^2}\right)[1-\cos(\varphi-
	\varphi_0)]+i\left(\frac{2c^2}{1+c^2}\right)\sin(\varphi-\varphi_0)\nn\\
	&=&
	1+(\sigma+\frac{3\mu}{4}+\frac{1}{2})[1-\cos(\varphi-\varphi_0)]
	+i\mu\sin(\varphi-\varphi_0)\;.
	\label{w222}
	\eea

Having obtained $\Gamma^2_\circ$ and $w^2_\circ$, we can use Eq.~(\ref{Pi-0})
to calculate the non-Abelian noncyclic geometric phase $\check\Gamma^2_\circ=
w^2_\circ\Gamma^2_\circ$. As seen from the above formulas the result will only 
depend on $\varphi_0$, $\varphi$, and $c=\cos\theta$. The expression for 
$\check\Gamma^2_\circ$ is rather lengthy. Therefore, we shall instead give
its trace $\Pi_\circ^2$ which is of physical importance,
	\bea
	\Pi^2_\circ &=& e^{i(\mu+\sigma)(\varphi-\varphi_0)/2}\left(
	{\cal X}\cos\left[\frac{\Delta(\varphi-\varphi_0)}{2}\right]+
	{\cal Y}\sin\left[\frac{\Delta(\varphi-\varphi_0)}{2}\right] \right)\;,~~~
	{\rm where}
	\label{trace-gp}\\
	{\cal X} &:=& \frac{1}{4}e^{-i(\varphi-\varphi_0)/2}[6+7\mu+4\sigma
	+(2-7\mu-4\sigma)\cos(\varphi-\varphi_0)+4i\sin(\varphi-\varphi_0)]
	,~~~{\rm and}\nn\\
	{\cal Y}&:=&\frac{i}{8\Delta}e^{-i(3\varphi-\varphi_0)/2}(
	e^{i\varphi_0}-e^{i\varphi})\left\{8\nu^2(1+e^{i(\varphi+\varphi_0)})
	+e^{i\varphi_0}[\mu(7\mu-5)-3(2+\mu)\sigma-4\sigma^2-2]+\right.\nn\\
	&&\left. e^{i\varphi}[\mu(9\mu-19)+(14-13\mu)\sigma+4\sigma^2+10]
	\right\}\;.\nn
	\eea
One can check that for $\varphi=\varphi_0$, $\Pi^2_\circ=2$, as expected. 

Furthermore, setting $\varphi=\varphi_0+2\pi$ in (\ref{trace-gp}), we obtained the 
trace of the cyclic non-Abelian geometric phase (\ref{G-0}) which is given by
	\be
	\left.\Pi^2_\circ\right|_{\rm cyclic}=-2 e^{i\pi(\mu+\sigma)}\cos(\pi\Delta)
	\label{cyclic-non-abelian-gp}
	\ee
In  this case, we can easily compute the eigenvalues of the non-Abelian cyclic 
geometric phase. They turn out to be $e^{\pm i \pi (\Delta+1)}$.

Now choosing  $\varphi_0=0$ and $c=\cos\theta=1/\sqrt{3}$, which are the value 
used in Tycko's experiment \cite{tycho}, we have $\Delta=\sqrt{889}/24\approx
1.24$ and
	\bea
	\Pi^2_\circ&\approx&\frac{1}{3}e^{0.10 i\varphi}
	\left\{(2+4\cos\varphi+3i\sin\varphi)\cos(0.62\varphi)+\right.\nn\\
	&&\left. 0.81[\cos(\varphi/2)+2.42 i\sin(\varphi/2)]\sin(\varphi/2)
	\sin(0.62\varphi)\right\}\;.\nn
	\eea

\section{Conclusion}

In this article we used the theory of dynamical invariants of Lewis and Riesefeld
to develop a general parameter space approach for the nonadiabatic geometric
phase. We introduced a set of time-dependent gauge-invariant geometric quantities
$\Pi^n(t)$ which we identified with the trace of certain time-dependent 
gauge-covariant geometric quantities $\check{\Gamma}^n(t)$. For a cyclic evolution 
with  period $T$,
$\check{\Gamma}^n(T)$ yields the non-Abelian cyclic geometric phase. In the Abelian
case, $\check{\Gamma}^n(t)/|\check{\Gamma}^n(t)|$ coincides with the Abelian
noncyclic geometric phase studied in the literature. Therefore, we identified the 
$\check{\Gamma}^n(t)$ as a non-Abelian noncyclic geometric phases factor. 

We also discussed the adiabatic limit $\check{\Gamma}^n_\circ(t)$ of 
$\check{\Gamma}^n(t)$. Again we showed that for the Abelian case
$\check{\Gamma}^n_\circ(t)/|\check{\Gamma}_\circ^n(t)|$ is the known adiabatic 
noncyclic geometric phase.

We have finally presented a through analysis of the Non-Abelian cyclic and
noncyclic geometric phases for a spin 1 quadrupole in a precessing field.

We wish to conclude this paper by pointing out the following observations.
	\begin{itemize}
	\item[1.] The original definition of the (nonadiabatic) non-Abelian cyclic
geometric phase is due to Anandan  \cite{anandan}. As pointed out in 
Ref.~\cite{jpa-98c} Anandan's definition is identical with the one given in terms
of the dynamical invariants. More specifically, the basis vectors 
$|\tilde\psi_a(t)\kt$ used by Anandan \cite{anandan} to yield the non-Abelian
connection one-form are precisely the basis eigenvectors $|\lambda_n,a;\theta(t)\kt$
of our approach. In fact, our approach may be viewed as a means to identify Anandan's
basis vectors $|\tilde\psi_a(t)\kt$ with the eigenvectors of a Hermitian operator, namely
a dynamical invariant $I[\theta(t)]$.
	\item[2.] In Refs.~\cite{pati,pati2}, Pati shows that the 
Abelian noncyclic geometric phase angle of Eq.~(\ref{pi}) may be written in
the form
	\be
	\check{\gamma}_n=\int_{\cal C} \Omega^n\;,
	\label{pati-form}
	\ee
where $\Omega^n=i\br\chi_n(t)|d|\chi_n(t)\kt$ and $|\chi_n(t)\kt$ is a properly 
scaled state vector. Although Pati terms $\Omega^n$ a connection one-form, 
he shows that indeed $\Omega^n$ is invariant under a gauge transformation.
This is because $\Omega$ is the difference of two connection one-forms, namely
the connection one-form $A^n$ and another connection one-form $P^n$ which also
depends on the initial state vector. As pointed out by one of the referees, it 
would be interesting to see whether Pati's results can be generalized to the 
non-Abelian case. Clearly, if one chooses a basis in the degeneracy subspace 
${\cal H}_{\lambda_n}(t)$ in which the non-Abelian noncyclic phase factor 
$\check{\Gamma}^n(t)$ is diagonal, then the diagonal elements may be 
treated as Abelian noncyclic phase factors and Pati's results may be used 
to express the corresponding phase angles in the form (\ref{pati-form})
provided that $\check{\Gamma}^n(t)$ (alternatively $w^n(t)$) is invertable.
In such a basis, $\check{\Gamma}^n={\rm diag}(g_1 e^{i\check{\gamma}_n^1},\cdots,
g_{l_n} e^{i\check{\gamma}_n^{l_n}})=:G^n e^{i S_n}$, where `diag$(\cdots)$' 
stands for a diagonal matrix with diagonal elements $\cdots$, $g_\ell$ and 
$\check{\gamma}_n^\ell$ are real, $G^n:={\rm diag}(g_1,\cdots,g_{l_n})$, and 
$S_n:={\rm diag}(\check{\gamma}_n^1,\cdots,\check{\gamma}_n^{l_n})$. In view 
of Pati's results, one may find an appropriate diagonal matrix of one-forms 
$\check{\Omega}^n$ such that $S_n=\int_{\cal C}\check{\Omega}^n$.
Furthermore, $\check{\Omega}^n$ will have the form $A^n-P^n$ for some
matrix of one-forms $P^n$.  Now one may postulate that $P^n$ is a connection 
one-form, so that under a gauge transformation $\check{\Omega}^n$ transforms 
covariantly. This together with the form of the diagonal elements of $P^n$ 
which is given by Pati \cite{pati2} are sufficient to obtain $\check{\Omega}^n$ 
in an arbitrary basis. A complete generalization of Pati's results to the
non-Abelian case would require a set of properly scaled state vectors
$|\chi_n,a;\theta\kt$ satisfying $\check{\Omega}^n_{ab}=i\br\chi_n,a;\theta|
d|\chi_n,b;\theta\kt$. The explicit form of the vectors $|\chi_n,a;\theta(t)\kt$
is not known to the author.
	\item[3.] The function $F$ introduced in section~7 may be used to relate the 
adiabatic and nonadiabatic Berry connection one-forms. Namely, the nonadiabatic 
connection one-form $A^n[\theta]$ is the pullback \cite{nakahara,jmp-94c} of the 
adiabatic connection one-form $A^n_\circ[R]$, provided that $F$ is
differentiable. This has been originally pointed out in Ref.~\cite{jpa-93}. 
However, in \cite{jpa-93} the existence of $F$ was assumed based on the 
evidence provided by the study of a magnetic dipole interacting with a precessing 
magnetic field. In the present article, we used the theory of dynamical invariants
to establish the existence of $F$ for a large class of quantum systems. In particular,
it is not difficult to see that they exist for the systems possessing a dynamical
group. This allows for the application of the holonomy interpretation of the
nonadiabatic geometric phase using a fiber bundle which has the parameter
space ${\cal M}$ of the invariant as its base space  \cite{jpa-93}. This bundle is 
the pullback bundle $F_*(\lambda)$ of the $U(\cun)$ bundle $\lambda$ used in 
the holonomy interpretation of the adiabatic geometric phase \cite{jpa-93,jmp-94c}. 
Note however that the function $F$ depends on the adiabaticity parameter 
\cite{pra-97a}. In general, there are certain values of the adiabaticity parameter for
which $F$ becomes ill-defined or discontinuous. The above constructions are valid only for
those values of the adiabaticity parameter for which $F$ is a differentiable function.
	 \item[4.] In our derivation of the noncyclic geometric phase, we assume that 
the Hamiltonian is a Hermitian operator. The generalization of our results to 
non-Hermitian Hamiltonians is straightforward. One needs the machinery 
of the non-Hermitian dynamical invariants and their biorthonormal eigenvectors to 
obtain a non-Hermitian analogue of the noncyclic geometric phase.
	\end{itemize}

\end{document}